\documentclass[aip,amsmath,amssymb,reprint]{revtex4-1}

\usepackage{graphicx}
\usepackage{dcolumn}
\usepackage{bm}
\usepackage{ulem}
\usepackage[utf8]{inputenc}
\usepackage[T1]{fontenc}
\usepackage{mathptmx}
\usepackage{etoolbox}
\usepackage{xcolor}
\makeatletter
\def\@email#1#2{
 \endgroup
 \patchcmd{\titleblock@produce}
  {\frontmatter@RRAPformat}
  {\frontmatter@RRAPformat{\produce@RRAP{*#1\href{mailto:#2}{#2}}}\frontmatter@RRAPformat}
  {}{}
}%
\makeatother
\begin{document}

\preprint{}

\title[Cryptocurrency inter-transaction time]{Analysis of inter-transaction time fluctuations in the cryptocurrency market.}

\author{Jarosław Kwapień}
\affiliation{
Complex Systems Theory Department, Institute of Nuclear Physics, Polish Academy of Sciences, Radzikowskiego 152, 31-342 Kraków, Poland}
\author{Marcin Wątorek}
\affiliation{Faculty of Computer Science and Telecommunications, Cracow University of Technology, ul.~Warszawska 24, 31-155 Krak\'ow, Poland}
\author{Marija Bezbradica} 
\affiliation{Adapt Centre, School of Computing, Dublin City University, Glasnevin, Dublin 9, Ireland}
\author{Martin Crane} 
\affiliation{Adapt Centre, School of Computing, Dublin City University, Glasnevin, Dublin 9, Ireland}
\email{martin.crane@dcu.ie}
\author{Tai Tan Mai} 
\affiliation{Adapt Centre, School of Computing, Dublin City University, Glasnevin, Dublin 9, Ireland}
\author{Stanisław Drożdż}
\affiliation{
Complex Systems Theory Department, Institute of Nuclear Physics, Polish Academy of Sciences, Radzikowskiego 152, 31-342 Kraków, Poland}
\affiliation{Faculty of Computer Science and Telecommunications, Cracow University of Technology, ul.~Warszawska 24, 31-155 Krak\'ow, Poland}
\email{jaroslaw.kwapien@ifj.edu.pl}

\date{\today}

\begin{abstract}
We analyse tick-by-tick data representing major cryptocurrencies traded on some different cryptocurrency trading platforms. We focus on such quantities like the inter-transaction times, the number of transactions in time unit, the traded volume, and volatility. We show that the inter-transaction times show long-range power-law autocorrelations. These lead to multifractality expressed by the right-side asymmetry of the singularity spectra $f(\alpha)$ indicating that the periods of increased market activity are characterised by richer multifractality compared to the periods of quiet market. We also show that neither the stretched exponential distribution nor the power-law-tail distribution are able to model universally the cumulative distribution functions of the quantities considered in this work. For each quantity, some data sets can be modeled by the former, some data sets by the latter, while both fail in other cases. An interesting, yet difficult to account for, observation is that parallel data sets from different trading platforms can show disparate statistical properties.
\end{abstract}

\maketitle

\begin{quotation}
The cryptocurrency market is the newest financial market that emerged from scratch not much more than a decade ago yet it has already managed to undergo a gradual shift from its infancy stage in 2012-2013 to an almost mature form at present. This shift has been observed primarily in the statistical properties of the cryptocurrency data, like the price return distributions, multifractal properties, trading activity, hedging opportunities, and so on~\cite{WatorekM-2021a,WatorekM-2021b,WatorekM-2021c}. Among many quantities characterising the cryptocurrency market, the inter-transaction time intervals, sometimes called waiting times~\cite{OswiecimkaP-2005a} or inter-trade durations~\cite{PolitiM-2008a}, and some related quantities like the number of transactions made in time unit still lack a proper analysis. Our work aims at filling this gap with an analysis based on high-quality data.
\end{quotation} 

\section{Introduction}

If compared with the price returns, the inter-transaction time (ITT) intervals $\delta t$ have relatively sparsely been a subject of research. A null-hypothesis approach might suggest that the transaction moments are uncorrelated and their generating mechanism is, thus, a Poisson-like process that produces an exponential pdf. The studies based on empirical data did not support this hypothesis, however. The inter-transaction times occur to be autocorrelated with long-range dependence~\cite{OswiecimkaP-2005a}. This observation was the basis for applying alternative models of the generating mechanism, like the autoregressive conditional duration processes~\cite{EngleR-1998a} and the continuous-time random walks~\cite{MontrollE-1965a,ScalasE-2000a,MainardiF-2000a,Kutner2003,Klamut2020,Klamut2021}. The probability that the next transaction will occur after at least a specific time $\tau$ can also be seen as no-trade survival probability $P(\tau)$. A natural distribution for such a quantity is the Weibull distribution:
\begin{equation}
P(x)=\alpha {x^{\alpha-1} \over x_0^{\alpha}} \exp {[-(x / x_0)^{\alpha}]}, \quad x > 0,
\end{equation}
where $\alpha > 0$.

The inter-transaction-time time series exhibit long-range power-law correlations.  This was shown, for instance, in~\cite{IvanovP-2004a} by applying the detrended fluctuation analysis (DFA) to ITT time series representing 30 large-cap American stocks. The time series were found to be persistent with two correlation regimes expressed by the Hurst exponents: $H=0.64\pm0.02$ for the intraday time scales and $H=0.94\pm0.05$ for the time scales longer than daily. This means that the ITT signals are long-range autocorrelated, which produces transaction clustering -- an effect that is analogous to volatility clustering~\cite{LiuY-1999a}. Transaction clustering is responsible for transaction frequency fluctuations and long-range autocorrelation of the number of transactions in time unit~\cite{PlerouV-2000a}. This significantly elevates the survival probability $P(\tau)$ for large $\tau$ with respect to uncorrelated time series and makes a special case of the Weibull distribution -- the stretched exponential (SE) distribution -- the appropriate model for describing the statistical properties of the ITT time series~\cite{LaherrereJ-1998a}. The cumulative distribution function (cdf) of SE has a particularly simple form
\begin{equation}
P_c(x) = \exp {[-(x/x_0)^{\alpha}]}, \quad 0 < \alpha \le 1.
\label{eq::stretched.cdf}
\end{equation}
For $\alpha=1$ it reduces to a standard exponential distribution.

Based on the tick-by-tick recordings representing the General Electric stock ITT in Oct 1999~\cite{RabertoM-2002a} and 30 stocks belonging to Dow Jones index over the same period~\cite{PolitiM-2008a}, it was shown that $P_c(x)$ can reproduce the empirical distributions for $0.73 \le \alpha \le 0.95$, depended on a stock. Mainardi et al.~\cite{MainardiF-2000a} analyzed ITTs for German bonds and found agreement between SE cdf and empirical data for small and medium-size ITTs with $\alpha\approx0.95$. A smaller $\alpha\approx0.82$ and 0.90 were reported for the Korean treasury bond futures market KOFEX~\cite{KimK-2003a}. The SE model was well fitted to the inter-transaction times of 30 American companies (1993-1996, $\delta t=1$). After a proper rescaling of the data, the empirical cdfs showed some degree of universality across different stocks with a mean value of the parameter $\alpha\approx0.72$~\cite{IvanovP-2004a}. The USD/JPY exchange rate 2002-2004 revealed much a smaller SE parameter $\alpha\approx0.59$~\cite{SazukaN-2007a}. A study involving the most liquid Chinese stocks in 2003 reported that $\alpha\approx 0.5$, ~\cite{RuanY-2011a}.

It is noteworthy that several data sets were modeled by the power-law distributions instead of SE. For example, ITT for the JPY/USD exchange rate between Oct 1998 and Mar 1999 were studied in~\cite{TakayasuH-2000a} and fitted by a power-law model with tail exponent $\gamma\approx1.8$ (i.e., in the L\'evy-stable domain). Moreover, a sample Chinese liquid stock (the years 2005-2006) and its warrant were found to differ from the SE model and be more power-law-like~\cite{RuanY-2011a}. The SE model fails to fit the data for the Forex exchange rates of major currencies if the waiting times between the best executable bid or ask prices are considered instead of the ITTs~\cite{ZhaoG-2013a}. A better choice is the log-normal distribution for a global fit covering both the small and large waiting times, but the best performance is observed if the power-law distribution is fitted to large waiting times and a certain variant of the El-Farol bar model is fitted to small waiting times~\cite{ZhaoG-2013a}. The $q$-Gaussian distributions~\cite{PolitiM-2008a,JiangZ-2008a} as well as the Mittag-Leffler function~\cite{SazukaN-2009a} were also reported to fit well to various ITT data. 

The stretched exponential function was also applied to model the return times when some quantity like, e.g., volatility, remains or recurs above certain threshold~\cite{WangF-2006a,QiuT-2008a,RenF-2009a,WangF-2009a,RenF-2009b,XieW-2014a}, the waiting times between certain extreme events~\cite{KaizojiT-2004a,SanthanamM-2008a,MengH-2012a}, the waiting times for a next price change~\cite{InoueJ-2010a}, the first-passage times~\cite{PerelloJ-2011a}, and the inter-order-cancellation intervals~\cite{GuG-2014a}. It is worth noting that for $\alpha > 1$ the SE function converts to the compressed exponential function, which sometimes has its application in finance (for instance, the over-exponential growth of the currency exchange rate in a case of heavy hyperinflation~\cite{MizunoT-2002a}).

\section{Data}

We consider a set of tick-by-tick time series of buy/sell transactions that were executed on several major trading platforms (Binance~\cite{Binance}, Bitfinex~\cite{Bitfinex}, Bitstamp~\cite{Bitstamp}, Coinbase~\cite{Coinbase}, HitBTC~\cite{HitBTC}, and Kraken~\cite{Kraken}) involving the most liquid cryptocurrencies: bitcoin (BTC), ether (ETH), ripple (XRP), and litecoin (LTC) expressed in USD (USDT in the case of Binance). These time series cover typically the period Jan 2017 -- Apr 2021, but the start and end dates may differ from that in individual cases. Tab.~\ref{tab::data.details} shows these dates for all time series along with the length $T$ of the series. We chose those 4 cryptocurrencies as they are the most liquid ones and the corresponding time series are the longest. From the raw data consisting of successive transactions that were concluded at time moments $t_i$ with $i=1,...,T$, we extract time series of inter-transaction intervals (or waiting times) $\delta t_i = t_{i+1}-t_i$. It should be noted that different trading platforms may have different recording precision and sometimes it happens that the time resolution is lower than the trading frequency, which leads to a situation in which ITT can be null. The lower the resolution is, the higher is the fraction $\chi$ of those $i$s for which $\delta t_i=0$. This effect can have a heavy impact on the statistical properties of the time series under study. Tab.~\ref{tab::data.statistics} collects the respective numbers. Apart from Kraken, the data from all the other platforms shows high fraction of zero ITTs ($16\% \le \chi \le 58\%$). The uniqueness of the Kraken data in this context comes from a fact that trading frequency on that platform is smaller than on the other platforms except for Bitstamp, so it is less likely that two transactions are less distant in time from each other than the temporal resolution of the data set. The case of Bitstamp is different as the temporal resolution of its data improved substantially during the period under study and in its early part that led to an excessive number of the null data points (and larger $\chi$) as compared to the later parts.

\begin{table}[]
\centering
\caption{Start and end dates of all time series considered in our study together with their length $T$ in data points.} \label{tab::data.details}
\scriptsize
\begin{tabular}{|c|c|c|c|c|c|c|}
\cline{2-7}
\multicolumn{1}{c|}{} & Bitfinex & Bitstamp & Coinbase & HitBTC & Kraken & Binance \\
\cline{2-7}
\cline{2-7}
\multicolumn{1}{c|}{} & \multicolumn{6}{c|}{Start date} \\ \hline
BTC & 1/1/2017 & 1/1/2017 & 1/1/2017 & 1/1/2017 & 1/1/2017 & 8/17/2017 \\ \hline
ETH & 1/1/2017 & 8/16/2017 & 1/1/2017 & 5/5/2017 & 1/1/2017 & 8/17/2017 \\ \hline
LTC & 1/1/2017 & 6/16/2017 & 1/1/2017 & 1/1/2017 & 1/1/2017 & 12/13/2017 \\ \hline
XRP & 5/19/2017 & 1/3/2017 & 2/26/2019 & 12/6/2017 & 5/18/2017 & 5/4/2018 \\ \hline
\multicolumn{1}{c|}{} & \multicolumn{6}{c|}{End date} \\ \hline
BTC & 4/6/2021 & 4/6/2021 & 4/6/2021 & 4/6/2021 & 4/6/2021 & 4/6/2021 \\ \hline
ETH & 4/6/2021 & 4/6/2021 & 4/6/2021 & 4/6/2021 & 4/6/2021 & 4/6/2021 \\ \hline
LTC & 4/6/2021 & 4/6/2021 & 4/6/2021 & 4/6/2021 & 4/6/2021 &4/6/2021 \\ \hline
XRP & 4/6/2021 & 4/6/2021 & 1/18/2021 & 4/6/2021 & 4/6/2021 &4/6/2021\\ \hline
\multicolumn{1}{c|}{} & \multicolumn{6}{c|}{\it T} \\ \hline
BTC & 116320238 & 39568217 & 138615908 & 72142694 & 30581727 & 739349486 \\ \hline
ETH & 56954769 & 12680067 & 94057512 & 33541056 & 22202050 & 342710847 \\ \hline
LTC & 23909764 & 5192202 & 58734675 & 15265870 & 5451199 & 126410087 \\ \hline
XRP & 36010677 & 17429435 & 26070113 & 9051495 & 6871839 & 169623206 \\ \hline
\end{tabular}
\end{table}

In order to put the results obtained for cryptocurrencies in perspective, we also consider two sets of ITT time series representing standard assets -- large cap stocks from the American and German stock markets: Bank of America (BAC), Johnson\&Johnson (JNJ), Intel (INT), J.P. Morgan (JPM), and Microsoft (MSFT), Deutsche Bank (DBK), DaimlerChrysler (DCX), Karstadt (KAR), Linde (LIN), Siemens (SIE), and Volkswagen (VOW). The corresponding time series cover the years 2010-2011 (American stocks) and 1998-1999 (German stocks).

\section{Results}

\subsection{Statistical properties over time}

Time series of ITTs are unsigned, thus their pdfs are asymmetric. In Tab.~\ref{tab::data.statistics} mean $\langle \delta t \rangle$ and standard deviation $\sigma_{\delta t}$ of the time series $\delta t_i$ are shown. Trading frequency varies both between the platforms and between the cryptocurrencies, with Binance and BTC being the most active, while Kraken, Bitstamp and LTC are the least active. Also there is high variability of $\delta t$ within the same platform and the same cryptocurrency -- in Tab.~\ref{tab::data.statistics} this is expressed by the significant values of $\sigma_{\delta t}$.

\begin{figure}
\includegraphics[width=0.45\textwidth]{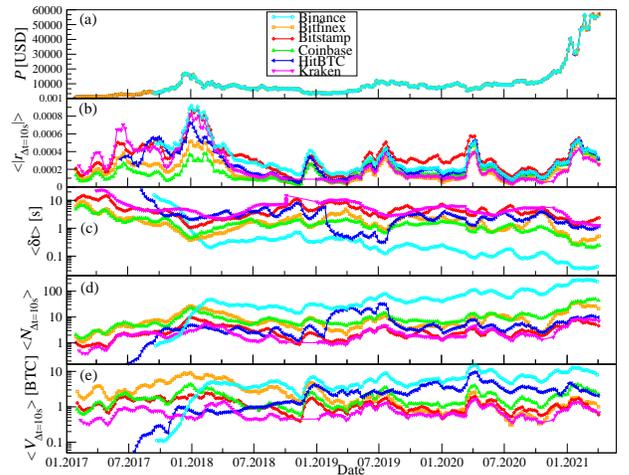}
\caption{A few basic statistics for the BTC inter-transaction interval time series calculated in 1-month-long moving window on different cryptocurrency trading platforms: Binance (cyan), Bitfinex (orange), Bitstamp (red), Coinbase (light green), HitBTC (blue), and Kraken (magenta): BTC price in USD (USDT in the case of Binance) (a), mean volatility $\langle |r_{\Delta t}| \rangle_t$ calculated in $\Delta t=10$s intervals (b), mean inter-transaction times $\langle \delta t \rangle_t$ (c), mean number of transactions $\langle N_{\Delta t} \rangle_t$ for $\Delta t=10$s (d), and mean volume traded $\langle V_{\Delta t} \rangle_t$ for $\Delta t=10$s (e).}
\label{fig::evolution.stats}
\end{figure}

The non-stationary character of $\delta t$ can be observed in Fig.~\ref{fig::evolution.stats}, where the evolution of a few statistical quantities calculated for the bitcoin data from different trading platforms is shown over the whole period 2017-2021 by using 1-month-long moving window. These quantities are: bitcoin price $p(t)$ expressed in USD, mean 10-second volatility $\langle |r_{\Delta t=10 {\rm s}}(t)| \rangle$, where $r_{\Delta t}(t)=\ln p(t+\Delta t)-\ln p(t)$ is logarithmic equal-time price return and $\Delta t$ is sampling time, mean ITT $\langle \delta t \rangle$, mean number of transactions $\langle N_{\Delta t}(t) \rangle$ and mean volume traded $\langle V_{\Delta t}(t) \rangle$ in intervals of length $\Delta t=10$s. All these quantities are non-stationary and their values and behavior change among the platforms. It is interesting to look at $\langle |r_{\Delta t}(t)|\rangle$, because the periods of its large values correspond to times of amplified market anxiety: the cryptocurrency bubble and subsequent crash in 2017-2018, the downward trend at the turn of 2018 and 2019, the transitional rally in mid-2019, the Covid-19 related panic in the early months of 2020, and start of another rally in the end of 2020. There is no qualitative difference in this pattern among the platforms, though their volatility levels are different.

Mean values of ITT for different platforms reflect differences in trading frequency on these platforms as well as developments in the cryptocurrency market, which attracts more and more investors as we approach the present day. The mean number of transactions in unit time $\langle N_{\Delta t} \rangle$ is, basically, an inverted mean ITT, thus the evolution of both quantities is closely coupled. The mean volume traded in unit time in principle depends on the number of transactions and prices, but it is also related to volatility: the larger $\langle r_{\Delta t}(t) \rangle$, the larger $\langle V_{\Delta t}(t) \rangle$. An interesting effect can be seen in Fig.~\ref{fig::evolution.stats} (bottom panel), where the largest volume was associated with the world's largest cryptocurrency trading platform -- Binance, which is natural. However, the second largest volume was associated with a platform of rather moderate size -- HitBTC, which is a counter-intuitive result especially if one observes that HitBTC was characterised by rather a small trading frequency. It looks as if on that platform a peculiar activity was taking place in 2019-2020 that involved a relatively small number of extremely large trades that were able to amplify volume up to a level comparable with the largest platforms. This can be related to wash-trading that may have taken place at that time~\cite{washtrading}.

\begin{table}[]
\centering
\caption{Basic statistics of the time series of inter-transaction times (ITTs): mean value $\langle \delta t \rangle$ and standard deviation $\sigma_{\delta t}$ (in seconds) of the inter-transaction times $\delta t$ and the fraction of null data points $\chi = (1/T) \times \#\{i\in[1,T]: \delta t_i=0\}$.}
\label{tab::data.statistics}
\scriptsize
\begin{tabular}{|c|c|c|c|c|c|c|}
\cline{2-7}
\multicolumn{1}{c|}{} & Bitfinex & Bitstamp & Coinbase & HitBTC & Kraken & Binance \\
\cline{2-7}
\cline{2-7}
\multicolumn{1}{c|}{} & \multicolumn{6}{c|}{$\langle \delta t \rangle$} \\ \hline
\multicolumn{1}{|l|}{BTC} & 1.158 & 3.405 & 0.972 & 1.867 & 4.395 & 0.153 \\ \hline
\multicolumn{1}{|l|}{ETH} & 2.363 & 9.064 & 1.431 & 3.692 & 6.063 & 0.329 \\ \hline
\multicolumn{1}{|l|}{LTC} & 5.629 & 23.153 & 2.292 & 8.815 & 24.691 & 0.820 \\ \hline
\multicolumn{1}{|l|}{XRP} & 3.404 & 7.708 & 2.294 & 11.624 & 17.852 & 0.539 \\ \hline
\multicolumn{1}{c|}{} & \multicolumn{6}{c|}{$\sigma_{\delta t}$} \\ \hline
\multicolumn{1}{|l|}{BTC} & 4.475 & 8.791 & 2.707 & 5.909 & 12.358 & 0.856 \\ \hline
\multicolumn{1}{|l|}{ETH} & 8.974 & 23.406 & 4.954 & 10.219 & 17.722 & 1.460 \\ \hline
\multicolumn{1}{|l|}{LTC} & 17.596 & 37.067 & 7.351 & 16.723 & 36.970 & 3.330 \\ \hline
\multicolumn{1}{|l|}{XRP} & 12.353 & 18.777 & 5.885 & 21.563 & 29.850 & 1.975 \\ \hline
\multicolumn{1}{c|}{} & \multicolumn{6}{c|}{$\chi$} \\ \hline
\multicolumn{1}{|l|}{BTC} & 0.575 & 0.453 & 0.373 & 0.451 & 0.00016 & 0.362 \\ \hline
\multicolumn{1}{|l|}{ETH} & 0.526 & 0.444 & 0.448 & 0.241 & 0.000104 & 0.342 \\ \hline
\multicolumn{1}{|l|}{LTC} & 0.495 & 0.369 & 0.470 & 0.169 & 0.000089 & 0.386 \\ \hline
\multicolumn{1}{|l|}{XRP} & 0.481 & 0.468 & 0.305 & 0.165 & 0.00011 & 0.349 \\ \hline
\end{tabular}
\end{table}

\subsection{Temporal correlations}

\begin{figure}
\includegraphics[width=0.45\textwidth]{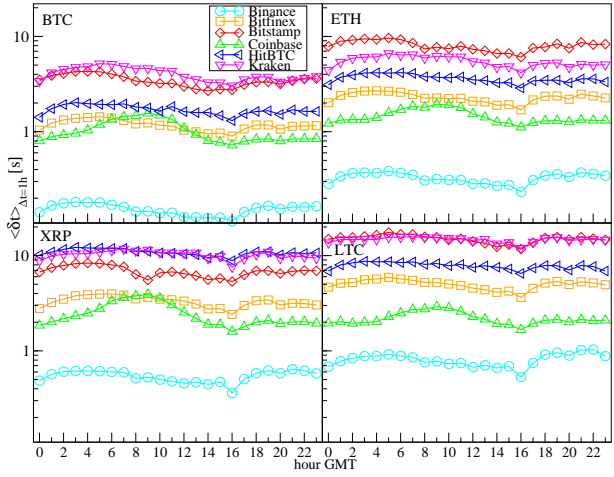}
\caption{Daily pattern of the mean inter-transaction time $\langle \delta t \rangle_{\Delta t}$ with $\Delta t=1$h calculated for different cryptocurrencies and different trading platforms.}
\label{fig::daily.pattern.crypto}
\end{figure}

\begin{figure}
\includegraphics[width=0.45\textwidth]{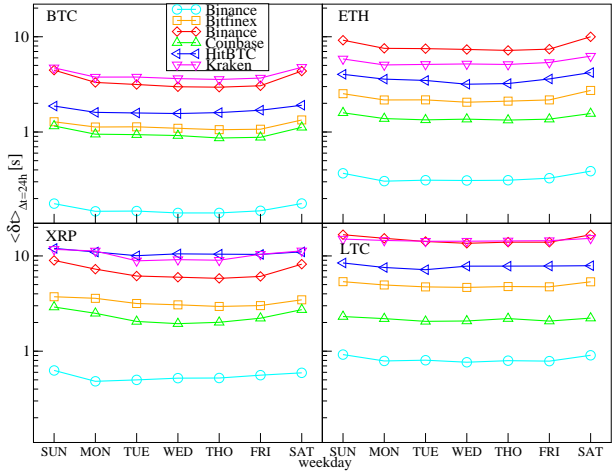}
\caption{Weekly pattern of the mean inter-transaction time $\langle \delta t \rangle_{\Delta t}$ with $\Delta t=24$h calculated for different cryptocurrencies and different trading platforms.}
\label{fig::weekly.pattern.crypto}
\end{figure}

\begin{figure}
\includegraphics[width=0.45\textwidth]{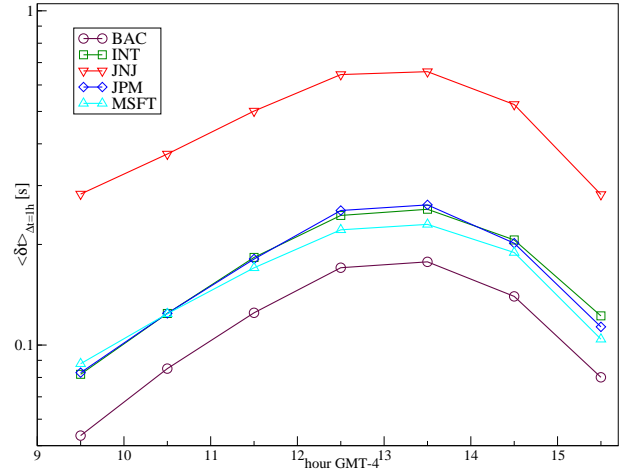}
\caption{Daily pattern of the mean inter-transaction time $\langle \delta t \rangle_{\Delta t}$ with $\Delta t=1$h calculated for sample large cap American stocks.}
\label{fig::daily.pattern.stocks}
\end{figure}

The existence of trends in $\langle \delta t \rangle_t$ and $\langle N_{\Delta t} \rangle_t$ suggest that long-range correlations are present in the corresponding time series. Such correlations can be detected in the simplest way by calculating the autocorrelation function
\begin{equation}
C(\tau_k)=(1/T_{\delta t}) {\langle \delta t_{i+k} \delta t_i \rangle_i \over \sigma_{\delta t}^2},
\label{eq::autocorrelation}
\end{equation}
where $\tau_k = k * \langle \delta t \rangle_t$. Because a well-known property of the inter-transaction times is seasonality, i.e. different parts of a trading day or week (e.g., Saturdays and Sundays in the case of cryptocurrencies) are characterised by different trading frequency~\cite{RabertoM-2002a}, before we apply Eq.~(\ref{eq::autocorrelation}), we have to eliminate this seasonality. Thus, we calculate mean ITT in 1-hour-long windows $\langle \delta t \rangle_{\Delta t=1{\rm h}}$ and obtain 24 data points for each trading day and then average them across all trading days in order to obtain a mean daily pattern of $\langle \delta t \rangle_{\Delta t=1{\rm h}}$. Daily patterns of ITT for cryptocurrencies (Fig.~\ref{fig::daily.pattern.crypto}) can be compared with the corresponding patterns for the American stocks (Fig.~\ref{fig::daily.pattern.stocks}). In the case of cryptocurrencies, we also remove weekly patterns by using the same procedure applied to the days of week (Fig.~\ref{fig::weekly.pattern.crypto}).

\begin{figure}
\includegraphics[width=0.45\textwidth]{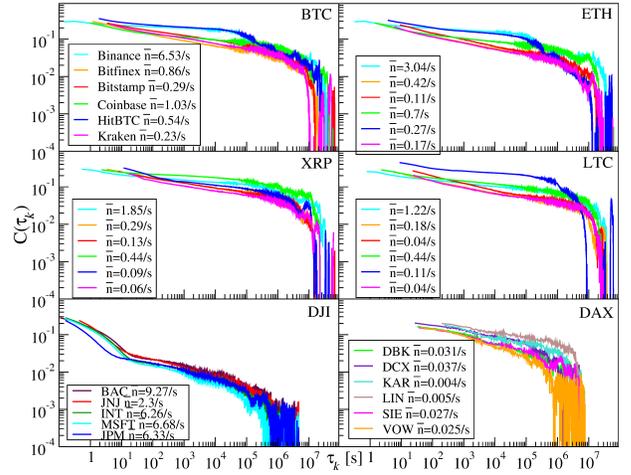}
\caption{Autocorrelation function $C(\tau_k)$ ($\tau_k = k * \langle \delta t \rangle_t$) of deseasonalised ITT time series $\delta t_i^{\rm des}$ calculated for the cryptocurrencies (BTC, ETH, XRP, and LTC) traded on different platforms (Binance, Bitfinex, Bitstamp, Coinbase, HitBTC, and Kraken), as well as sample American stocks from Dow Jones (BAC, INT, JNJ, JPM, MSFT) and sample German stocks from DAX30 (DBK, DCX, KAR, LIN, SIE, VOW). Mean trading frequency $\overline{n}=1/\langle\delta t\rangle_t$ of each instrument is given in the legend boxes.}
\label{fig::autocorrelation}
\end{figure}

Next, each value of $\delta t_i$ is divided by the corresponding pattern value and we get a time series of deseasonalised ITTs $\delta t_i^{\rm des}$. Now we can compute the autocorrelation function (ACF), $C(\tau_k)$ -- Fig.~\ref{fig::autocorrelation}. All financial assets considered in this study show clear long-range autocorrelation (with even a power-law decay in some cases) over 7 orders of magnitude up to $10^7$s (almost 4 months) for the cryptocurrencies and $10^6$s (2 weeks) for the stocks. A few additional observations can be made. While the memory effects last equally long for the American and German stocks, for the latter they are stronger in magnitude (larger values of $C(\tau_k)$). For the cryptocurrencies, the autocorrelation magnitude is between these two cases. Except for HitBTC that shows distinct behaviour of $C(\tau_k)$ than the other platforms, the autocorrelation range expressed in real-time units for a given cryptocurrency does not depend significantly on a platform even though the trading frequency differs among the platforms. We may conclude that the memory effects must thus be related to some external factors other than the inner dynamics of trading on these platforms.

A power-law decay of the autocorrelation function is observed for the processes that show multiscaling~\cite{KwapienJ-2012a}, so it is natural to ask whether the same can be observed in the present case as broad singularity spectra that indicate rich multifractality have already been reported in literature in a context of the ITT time series representing stocks~\cite{OswiecimkaP-2005a,JiangZ-2009a,RuanY-2011a,JiangZ-2019a}. A convenient way to characterise fractal properties of time series is by using multifractal detrended fluctuation analysis (MFDFA)~\cite{KantelhardtJ-2002a}. 

We start from a time series of ITTs $\delta t_i^{\rm des}$, $i=1,...,T$. For a given temporal scale $s$ ($s_{\rm min} \le s \le s_{\rm max}$), we divide the time series into segments of length $s$ by starting from both ends and obtaining $M_s=2 \lfloor{N/s} \rfloor$ segments total ($\lfloor\cdot\rfloor$ means integer part here). In each segment $\nu$ ($\nu=1,...,M_s$), we integrate data points and construct a detrended signal profile $\Delta_i(s,\nu)$:
\begin{equation}
\Delta_i (s,\nu) = \sum_{j=1}^i \delta t_{j+s\nu}^{\rm des} - P_{\nu}^{(m)}(i),
\label{eq::mfdfa.signal.profiles}
\end{equation}
where $P_{\nu}^{(m)}(i)$ is an $m$th-degree polynomial. Next we calculate the detrended variance for each segment:
\begin{equation}
f^2(s,\nu) = {1 \over s} \sum_{i=1}^s \Delta_i^2(s,\nu)
\label{eq::mfdfa.viariance}
\end{equation}
and use the so-defined variances to compute a family of fluctuation functions $F_q(s)$ of order $q$:
\begin{equation}
F_q(s) = \Big\{ {1 \over M_s} \sum_{\nu=1}^{M_s} f^2(s,\nu)^{q/2} \Big\}^{1/q}.
\label{eq::mfdfa.fluctuation.function}
\end{equation}
Fluctuation functions are calculated independently for a range of different scales $s$ with a typical $s_{\rm min}$ larger than the longest sequence of zeros in the original time series $\delta t_i^{\rm des}$ and $s_{\rm max}$ equal to $T/10$~\cite{OswiecimkaP-2006a}. Fractal signals give a power-law dependence on scale:
\begin{equation}
F_q(s) \sim s^{h(q)},
\label{eq::hurst}
\end{equation}
where a family of the generalized Hurst exponents $h(q)$ allow for distinguishing mono- and multifractal type of scaling: if $h(q) \neq {\rm const}$, the time series is multifractal, while it is monofractal otherwise. Another quantity that is able to distinguish both types of fractality is singularity spectrum $f(\alpha)$:
\begin{equation}
\alpha(q) = h(q) + q h'(q), \quad f(\alpha) = q (\alpha(q) - h(q)) + 1,
\label{eq::singularity}
\end{equation}
where $h'(q)$ denotes the first derivative of $h(q)$ with respect to $q$.

\begin{figure}
\includegraphics[width=0.45\textwidth]{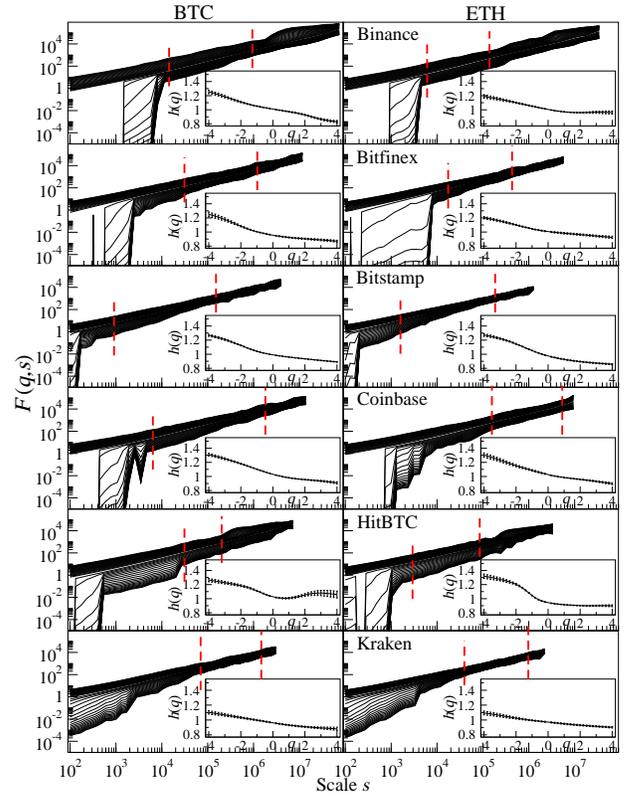}
\caption{(Main plots) Fluctuation function $F_q(s)$ calculated for the deseasonalised inter-transaction-time time series $\delta t_i^{\rm des}$ representing BTC (left) and ETH (right) and different trading platforms (top to bottom). The upper cut-off scale $s_{\rm max}$ depends on time series length and $q\in[-4,4]$. (Insets) The generalized Hurst exponents $h(q)$ estimated from the range of $s$ indicated by dashed red lines.}
\label{fig::fluctuation.function.BTC.ETH}
\end{figure}

\begin{figure}
\includegraphics[width=0.45\textwidth]{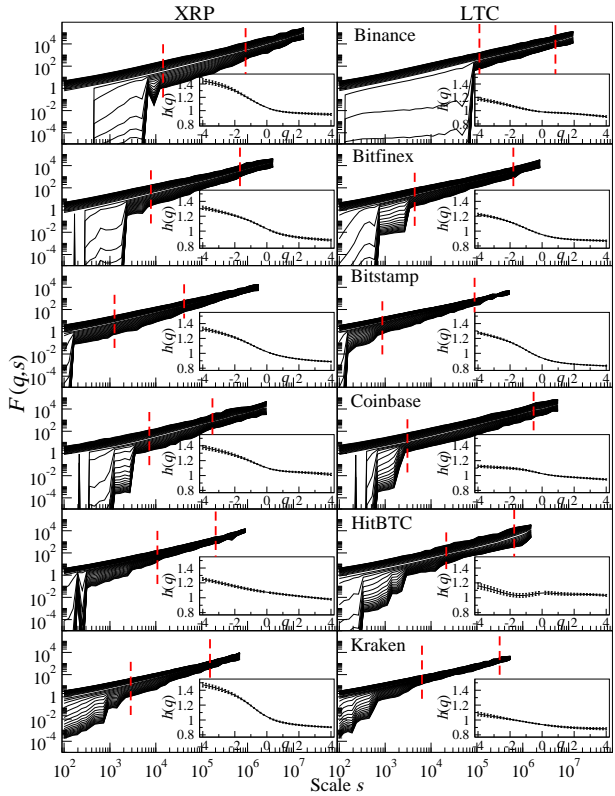}
\caption{(Main plots) Fluctuation function $F_q(s)$ calculated for the deseasonalised inter-transaction-time time series $\delta t_i^{\rm des}$ representing XRP (left) and LTC (right) and different trading platforms (top to bottom). The upper cut-off scale $s_{\rm max}$ depends on time series length and $q\in[-4,4]$. (Insets) The generalized Hurst exponents $h(q)$ estimated from the range of $s$ indicated by dashed red lines.}
\label{fig::fluctuation.function.XRP.LTC}
\end{figure}

We calculate fluctuation functions $F_q(s)$ for each deseasonalised ITT time series from out data sets. We choose $s_{\rm min}$ and $s_{\rm max}$ according to the above rules for each time series individually and fix the R\'enyi-like parameter $q$ to be within [-4,4]. The results obtained for the cryptocurrencies are collected in the main plots of Fig.~\ref{fig::fluctuation.function.BTC.ETH} (BTC and ETH) and Fig.~\ref{fig::fluctuation.function.XRP.LTC} (XRP and LTC). Except for the shortest time scales, for which the overall behaviour of $F_q(s)$ becomes heavily distorted especially for $q<0$, due to a significant number of the intervals with $\delta t_i^{\rm des}=0$, the fluctuation functions show fractal (parallel lines) and sometimes multifractal scaling (spread lines). This result is supported by the plots of the generalized Hurst exponent $h(q)$, which in each case is, roughly, a monotonically decreasing function of $q$ (insets of Fig.~\ref{fig::fluctuation.function.BTC.ETH} and Fig.~\ref{fig::fluctuation.function.XRP.LTC}).

\begin{figure}
\includegraphics[width=0.45\textwidth]{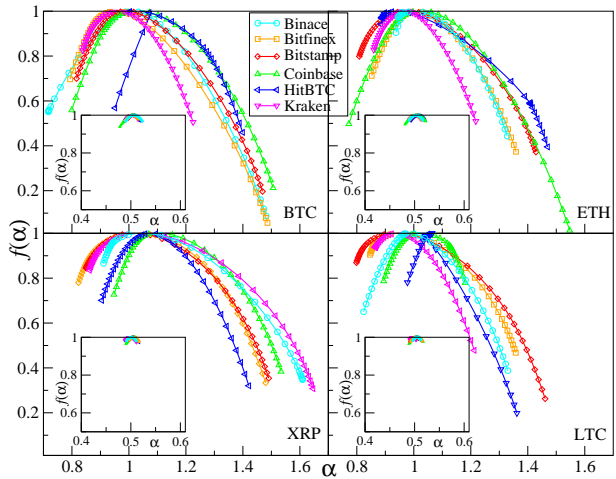}
\caption{Singularity spectra $f(\alpha)$ calculated for the deseasonalised inter-transaction-time time series $\delta t_i^{\rm des}$ representing cryptocurrencies: BTC (top left), ETH (top right), XRP (bottom left), and LTC (bottom right) as well as different trading platforms: Binance, Bitfinex, Bitstamp, Coinbase, HitBTC, and Kraken. The original time series (main plots) are compared with their shuffled surrogates (insets).}
\label{fig::singularity.spectra.crypto}
\end{figure}

In order to visualise it in a more straightforward manner, we calculate the singularity spectra $f(\alpha)$ for each function $F_q(s)$ that is power-law over at least two decades of $s$. These spectra are shown in Fig.~\ref{fig::singularity.spectra.crypto}.

What draws immediate attention is the broad shape of the presented spectra with no monofractal, point-like case regardless of the asset and the platform. Moreover, all the spectra show a significant right-side asymmetry, which means that the multifractal behaviour is observed primarily among small $\delta t_i^{\rm des}$, while for large ITTs the multifractality is less prominent~\cite{DrozdzS-2015a}. Such observation indicates that in periods of more intensive market activity that corresponds to small values of $\delta t_i$ the multifractality becomes richer than during the less intensive trading. As the opposite type of asymmetry, i.e., the left-hand side one, is typically observed in time series of the price returns, the right-hand side asymmetry has been expected to occur here for ITTs (small $\langle \delta t^{\rm des} \rangle$ leads to large $N_{\Delta t}$ and it in turn leads to large $r_{\Delta t}$). It is also worth noting that the time series, which we analyse here, are long enough that the effects of spurious broadening of $f(\alpha)$ that are common in short signals~\cite{DrozdzS-2009a,ZhouWX-2012a} with heavy-tailed pdfs disappear here. In order to show it, we created the shuffled surrogates of the original ITT signals and calculated the corresponding singularity spectra. They are displayed in Fig.~\ref{fig::singularity.spectra.crypto} for the four cryptocurrencies (insets). Their width is much smaller than in the case of the original time series and suggests that they are monofractal. Therefore we feel free to conclude that the multifractality is caused by the long-range autocorrelations in time series of ITTs.

Looking at more detail, in the main plots of Fig.~\ref{fig::singularity.spectra.crypto}, we see that there is a difference in the widths of the spectra  among the cryptocurrencies with, typically, the widest being the ones for ripple (XRP) and the narrowest being the ones for litecoin (LTC). There is no regularity, however, in the widths of the spectra associated with a particular cryptocurrency among different trading platforms.

\begin{figure}
\includegraphics[width=0.45\textwidth]{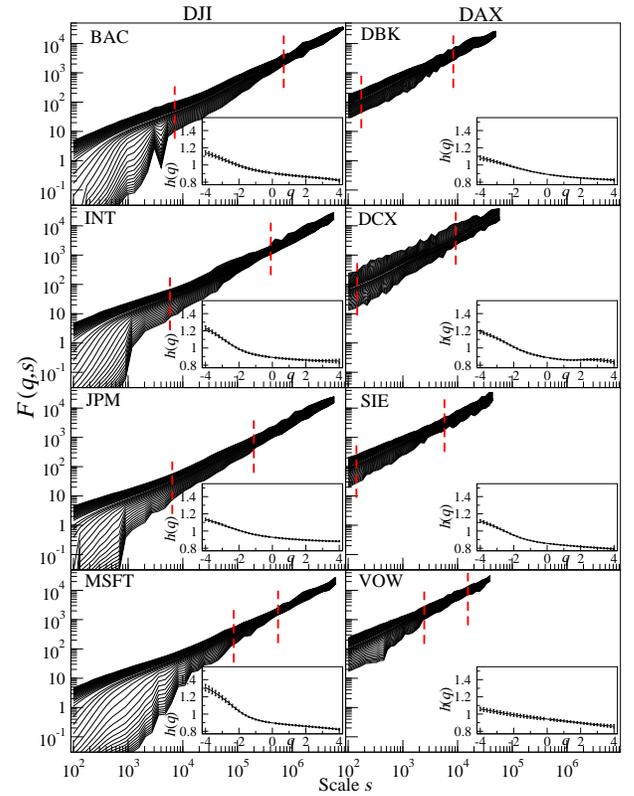}
\caption{(Main plots) Fluctuation function $F_q(s)$ calculated for the deseasonalised inter-transaction-time time series $\delta t_i^{\rm des}$ representing sample American (left) and German (right) large cap stocks (top to bottom). The cut-off scales depend on time series length. (Insets) The generalized Hurst exponents $h(q)$ estimated from the range of $s$ indicated by dashed red lines.}
\label{fig::fluctuation.function.stocks}
\end{figure}

\begin{figure}
\includegraphics[width=0.45\textwidth]{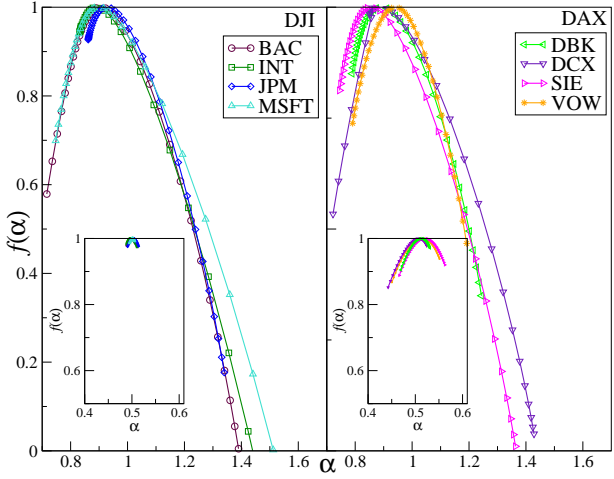}
\caption{Singularity spectra $f(\alpha)$ calculated for the deseasonalised inter-transaction-time time series $\delta t_i^{\rm des}$ representing sample large cap stocks from the American (left) and German (right) markets. The original time series (main plots) are compared with their shuffled surrogates (insets).}
\label{fig::singularity.spectra.stocks}
\end{figure}

Fluctuation functions calculated for the stocks are shown in Fig.~\ref{fig::fluctuation.function.stocks} and the corresponding singularity spectra in Fig.~\ref{fig::singularity.spectra.stocks}. The functions $F_q(s)$ for the American stocks reveal a power-law scaling over a longer range of scales than the German stocks do, which can be explained by the longer time series considered in the former case (not only the German stock market is smaller and less liquid than the American one, but also the data from the American market were 12 years later, when the trading frequency was much larger $:\langle \delta t^{\textrm{US}} \rangle\approx0.2$s and $\langle \delta t^{\textrm{GER}} \rangle\approx33$s). The spectra $f(\alpha)$ exhibit a right-side asymmetry as is the case for the cryptocurrencies -- see Fig.~\ref{fig::singularity.spectra.stocks}. Even though these spectra are broader for the surrogate time series of the German stocks than for the American ones (see insets in Fig.~\ref{fig::singularity.spectra.stocks}), we conclude that in both cases the surrogates may be considered as monofractal (the ITT time series of the German stocks are significantly shorter than the time series for their American counterparts because of the less intensive trading activity in Frankfurt in 1998-1999 than in New York in 2010-2011).

\begin{figure}
\includegraphics[width=0.45\textwidth]{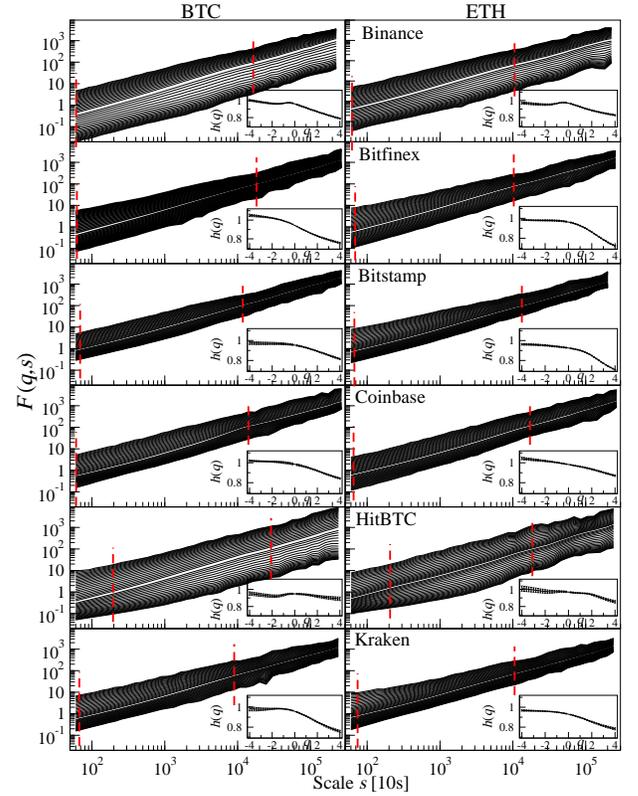}
\caption{(Main plots) Fluctuation function $F_q(s)$ calculated for the deseasonalised number of transactions $N_{\Delta t}^{\rm des}$ in intervals of $\Delta t=10$s for BTC (left) and ETH (right) and for different trading platforms (top to bottom). The upper cut-off scale $s_{\rm max}$ is the same in each case and $q\in[-4,4]$. (Insets) The generalized Hurst exponents $h(q)$ estimated from the range of $s$ indicated by dashed red lines.}
\label{fig::number.fluctuation.function.BTC.ETH}
\end{figure}

\begin{figure}
\includegraphics[width=0.45\textwidth]{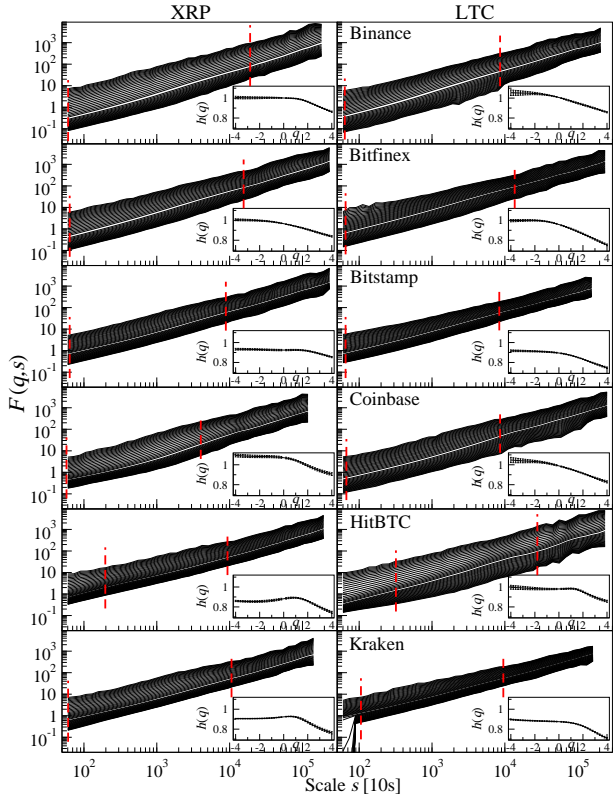}
\caption{(Main plots) Fluctuation function $F_q(s)$ calculated for the deseasonalised number of transactions $N_{\Delta t}^{\rm des}$ in intervals of $\Delta t=10$s for XRP (left) and LTC (right) and for different trading platforms (top to bottom). The upper cut-off scale $s_{\rm max}$ is the same in each case and $q\in[-4,4]$. (Insets) The generalized Hurst exponents $h(q)$ estimated from the range of $s$ indicated by dashed red lines.}
\label{fig::number.fluctuation.function.XRP.LTC}
\end{figure}

Each time series of ITT include a large number of null values $\delta t_i=0$ since temporal resolution of the recorded trade data can be worse than the actual inter-transaction intervals. Some of our data sets suffer from such a problem. This is why, in order to eliminate the effect of null data on the multifractal analysis, we consider time series of the number of transactions in time interval of $\Delta t=10$s: $N_{\Delta t,i}$, which comprises far fewer zeros. However, before we perform MFDFA on this quantity, we must note that $N_{\Delta t,i}$ undergoes seasonal fluctuations similar to those of $\langle \delta t_i \rangle_{\Delta t}$. It is therefore recommended to remove this seasonality in the same manner as it was done with $\delta t_i$ above and to focus on $N_{\Delta t,i}^{\rm des}$ henceforth. Fig.~\ref{fig::number.fluctuation.function.BTC.ETH} and Fig.~\ref{fig::number.fluctuation.function.XRP.LTC} present the corresponding fluctuation functions calculated for the cryptocurrencies. Indeed, the plots of $F_q(s)$ are now much more tame and homogeneous throughout the scales. Since the length of all time series is now comparable (see the dates in Tab.~\ref{tab::data.details}), we may fix $s_{\rm min}$ and $s_{\rm max}$.

\begin{figure}
\includegraphics[width=0.45\textwidth]{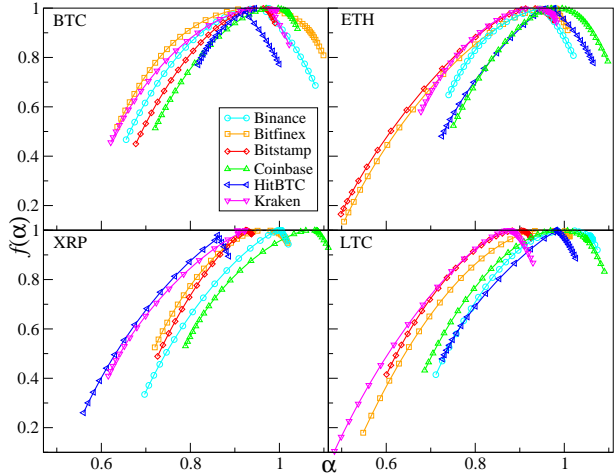}
\caption{Singularity spectra $f(\alpha)$ calculated for the deseasonalised number of transactions $N_{\Delta t}^{\rm des}$ in intervals of $\Delta t=10$s for the cryptocurrencies: BTC (top left), ETH (top right), XRP (bottom left), and LTC (bottom right) as well as different trading platforms: Binance, Bitfinex, Bitstamp, Coinbase, HitBTC, and Kraken.}
\label{fig::number.singularity.spectra}
\end{figure}

Unlike the case of ITT time series, the singularity spectra in Fig.~\ref{fig::number.singularity.spectra} reveal now a left-side asymmetry. Such a shape of $f(\alpha)$ provides us with information about the more visible multifractal properties of large fluctuations of $N_{\Delta t,i}^{\rm des}$ and much suppressed multifractality of small fluctuations~\cite{DrozdzS-2015a}. However, since the following relation holds $N_{\Delta t,i} = \langle \delta t_i \rangle_{\Delta t}^{-1}$, it becomes obvious why the left-side asymmetry of $f(\alpha)$ in Fig.~\ref{fig::singularity.spectra.crypto} transforms here into the right-side asymmetry.

The MFDFA formalism given by Eqs.~(\ref{eq::mfdfa.signal.profiles})-(\ref{eq::mfdfa.fluctuation.function}) can easily be generalized to two cross-correlated time series $x_i$ and $y_i$. The detrended covariance is then defined as
\begin{equation}
f_{\rm XY}^2(s,\nu) = {1 \over s} \sum_{i=1}^s X_i(s,\nu) Y_i(s,\nu),
\label{eq::mfdcca.covariance}
\end{equation}
where $X_i$ and $Y_i$ are signal profiles (Eq.~(\ref{eq::mfdfa.signal.profiles})) of $x_i$ and $y_i$, respectively.
The fluctuation functions are now given by
\begin{equation}
F_q^{\rm XY}(s) = \Big\{ {1 \over M_s} \sum_{\nu=1}^{M_s} {\rm sign}(f_{\rm XY}^2(s,\nu) |f_{\rm XY}^2(s,\nu)|^{q/2} \Big\}^{1/q},
\label{eq::mfdcca.fluctuation.function}
\end{equation}
where the sign function enters the formula in order to ensure continuity of $F_q^{\rm XY}(s)$, while the modulus prevents it from being complex. For X=Y the definition~(\ref{eq::mfdcca.fluctuation.function}) reduces to Eq.~(\ref{eq::mfdfa.fluctuation.function}). Having defined the fluctuation functions for $F_q^{\rm XY}(s)$, $F_q^{\rm XX}(s)$, and $F_q^{\rm YY}(s)$, we can calculate the $q$-dependent detrended cross-correlation coefficient $\rho_q(s)$, which is a counterpart of the Pearson cross-correlation coefficient for non-stationary signals~\cite{KwapienJ-2015a}:
\begin{equation}
\rho_q(s) = {F_q^{\rm XY}(s) \over \sqrt {F_q^{\rm XX}(s) F_q^{\rm YY}(s)}},
\end{equation}
which satisfied the relation $-1 \le \rho_q(s) \le 1$ for $q>0$. In this case, for independent time series $\rho_q(s)=0$, for perfectly correlated time series $\rho_q(s)=1$, and for perfectly anti-correlated time series $\rho_q(s)=-1$ for all scales $s$. For $q\le 0$ the coefficient $\rho_q(s)$ can assume values beyond the interval [-1,1] and its interpretation requires more a subtle approach~\cite{KwapienJ-2015a}. 

\begin{figure}
\includegraphics[width=0.45\textwidth]{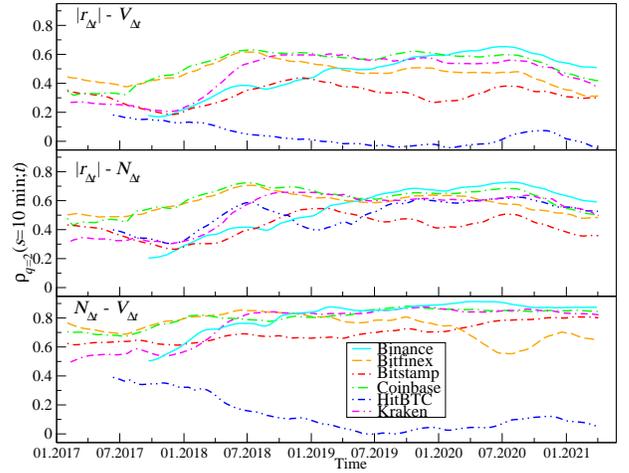}
\caption{Evolution of the $q$-dependent detrended cross-correlation coefficient $\rho_{q=2}(s=10\textrm{min};t)$ calculated over 1-month-long moving window for the deseasonalised time series of the number of transactions $N_{\Delta t}^{\rm des}$, volatility $|r_{\Delta t}^{\rm des}|$, and volume traded $V_{\Delta t}^{\rm des}$ of BTC in intervals of $\Delta t=10$s. Evolution of $\rho_{q=2}(s=10\textrm{min};t)$ for volatility and volume (top), volatility and the number of transactions (middle), and the number of transactions and volume (bottom) is shown for data from different trading platforms: Binance, Bitfinex, Bitstamp, Coinbase, HitBTC, and Kraken.}
\label{fig::rho.BTC}
\end{figure}

For classical financial markets, such quantities as transaction frequency, the number of transactions in some interval, volume traded, and volatility are related either statistically, causally, or both. For instance, it was shown in~\cite{GopikrishnanP-2000a,PlerouV-2000a} based on data from the American stock markets that fluctuations in the number of transactions in a time interval $N_{\Delta t}$ have a strong impact on the traded volume. The inter-transaction times and price returns were also reported to be correlated~\cite{RepetowiczP-2004a,MeerschaertM-2006a}. Also in the case of bitcoin correlations between volatility and volume were studied~\cite{BTC2018}. It is thus interesting to look at the respective statistical relations also in the case of the cryptocurrency market. We use the deseasonalised time series of the number of transactions $N_{\Delta t}^{\rm des}$, volatility $|r_{\Delta t}^{\rm des}|$, and volume $V_{\Delta t}^{\rm des}$ for $\Delta t=10$s and compute the coefficients $\rho_q(s)$ for all pairs of these time series. The results obtained for the BTC time series from different trading platforms are presented in Fig.~\ref{fig::rho.BTC}.

Although all three pairs of the considered quantities were substantially cross-correlated throughout the years 2017-2021 on all the platforms besides HitBTC, their cross-correlations were relatively small in 2017 and then gradually increased in 2018 and remained strong afterwards (with some moderate fluctuations in some cases). On average, $V_{\Delta t}^{\rm des}$ and $N_{\Delta t}^{\rm des}$ were the strongest correlated, next were the correlations between $|r_{\Delta t}^{\rm des}|$ and $N_{\Delta t}^{\rm des}$, while $V_{\Delta t}^{\rm des}$ and $|r_{\Delta t}^{\rm des}|$ were correlated the weakest (Fig.~\ref{fig::rho.BTC}). Among the platforms, Bitfinex and Coinbase showed the strongest correlations in 2017 and 2018, while Coinbase and Binance dominated from 2019 to 2021. Bitstamp showed a sizeably smaller level of $\rho_{q=2}(s=\textrm{10min},t)$ for the whole period, but nevertheless it was still significant. A notable exception in this picture is the platform HitBTC, where volume used to exhibit a moderate level of cross-correlation with volatility (top panel in Fig.~\ref{fig::rho.BTC}) and the number of transactions (bottom panel) in 2017, but this changed in 2018 when $\rho_q(s;t)$ started to decrease gradually to very low levels below 0.1 (for $N_{\Delta t}^{\rm des}$) and even to almost 0 (for $|r_{\Delta t}^{\rm des}|$) in mid-2019. This trend reversed to some degree in 2020, but even then $\rho_q=2(s=10\textrm{min};t)$ remained below 0.2 at its maximum at the turn of 2020 and 2021. Only the relation between volatility and the number of transactions was resembling partially its counterparts from the other platforms. The cross-correlation peculiarities observed in the case of HitBTC can also be explained by the wash-trading mechanism~\cite{Wash2021,Wash2022}.

\subsection{Inter-transaction time and volume distributions}

The price return and volatility fluctuations in time series from the financial markets have been a subject of extensive studies since the very beginnings of econometrics~\cite{BachelierL-1900a,MandelbrotB-1963a,MantegnaR-1995a,LiuY-1999a,GopikrishnanP-1999a,PlerouV-1999a,DrozdzS-2003b,DrozdzS-2007a,WatorekM-2021a}. Although there is no consensus among the researchers, which of the statistical distributions describes the pdf tail behaviour of such fluctuations with a satisfactory approximation, a substantial body of evidence indicates that one of the most accurate candidate distributions is the power-law pdf in its standard and exponentially truncated variants with the tail exponent $\beta\approx 3$~\cite{MantegnaR-1995a,PlerouV-1999a,GopikrishnanP-1999a}.

\begin{figure}
\includegraphics[width=0.45\textwidth]{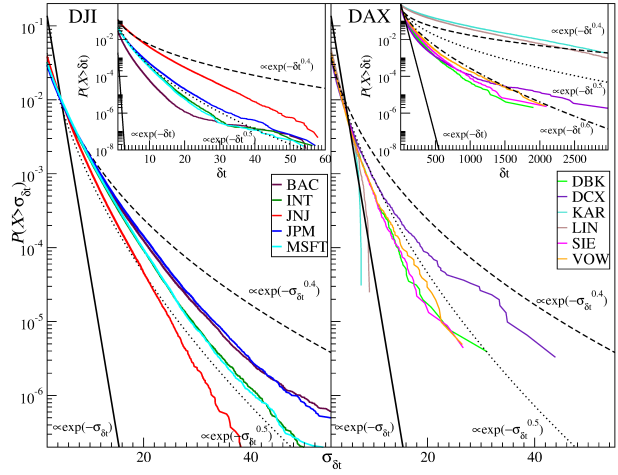}
\caption{Cumulative distribution functions of the normalized inter-transaction times $\hat{\delta t_i}$ in units of standard deviation $\sigma_{\delta t}$ (main plots) and the original inter-transaction times $\delta t_i$ in seconds (insets) for sample large cap stocks from the American (left) and German stock markets (right). Different variants of the SE model with $\alpha=0.4$ (dashed), $\alpha=0.5$ (dotted), $\alpha=0.6$ (dash-dotted), and $\alpha=1.0$ (solid) serve as guide for an eye.}
\label{fig::cdf.ITT.stocks}
\end{figure}

\begin{figure}
\includegraphics[width=0.45\textwidth]{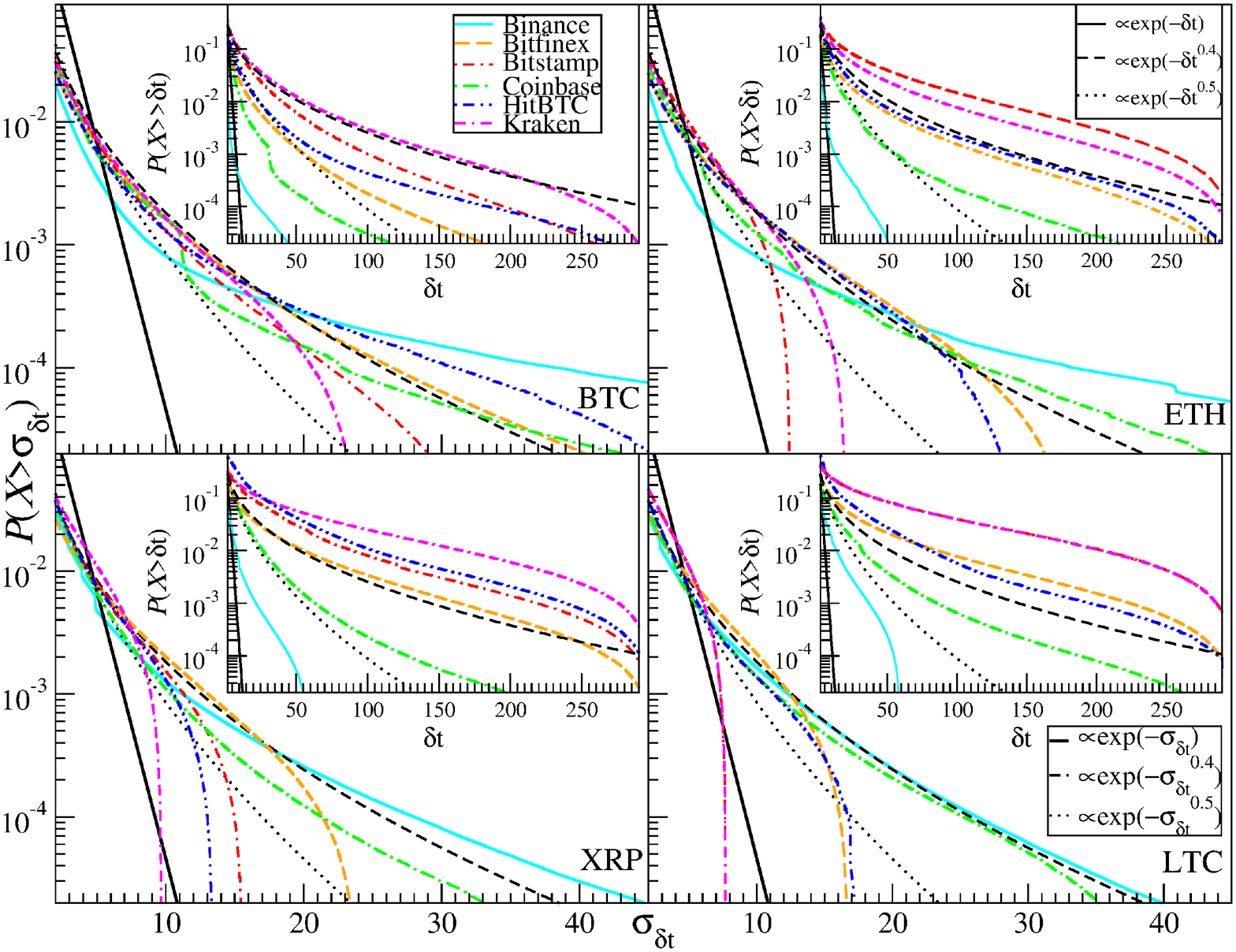}
\caption{Cumulative distribution functions of the normalized inter-transaction times $\hat{\delta t_i}$ in units of standard deviation $\sigma_{\delta t}$ (main plots) and the original inter-transaction times $\delta t_i$ in seconds (insets) for the cryptocurrencies: BTC (top left), ETH (top right), XRP (bottom left), and LTC (bottom right). In each plot cdfs for the data from six trading platforms are shown: Binance, Bitfinex, Bitstamp, Coinbase, HitBTC, and Kraken. Different variants of the SE model with $\alpha=0.4$ (dashed), $\alpha=0.5$ (dotted), and $\alpha=1.0$ (solid) serve as guide for an eye.}
\label{fig::cdf.ITT.crypto}
\end{figure}

Now we focus our attention on the statistical properties of the time series representing the quantities considered in our study. We start from calculating the cdfs of ITTs $\delta t_i$ that are then plotted in the insets in Fig.~\ref{fig::cdf.ITT.crypto}. According to earlier studies reported in literature that analyzed the ITT time series from the classical financial markets, such series can be modelled by various statistical distributions, including the stretched exponentials~\cite{MainardiF-2000a,RabertoM-2002a,KimK-2003a,IvanovP-2004a,SazukaN-2007a,PolitiM-2008a}, the power-laws~\cite{TakayasuH-2000a,RuanY-2011a}, and some other functions~\cite{JiangZ-2008a,ZhaoG-2013a}. Before we show the results for the cryptocurrencies, Fig.~\ref{fig::cdf.ITT.stocks} shows the ITT cdfs for the time series representing the American and German stocks. The plots use semi-logarithmic scale that allows us to distinguish a standard exponential distribution (which is a straight line in this case) from the SE distributions (nonlinear curves). It is evident from the plots that the exponential model may be discarded for the American stocks and for a majority of the German large cap ones, but the less capitalized stocks - KAR and LIN - traded on the German market can be close to this case (which suggests that the transactions involving such stocks may show little autocorrelation). The SE model performs much better overall and its specific realisations can approximate the empirical data, especially for $\alpha > 0.4$. The cdfs in the main panels were created for the normalized time series in order to make the distributions comparable with each other. However, they do not allow for a direct assessment of how large are the largest data points in absolute sense. This is why insets in Fig.~\ref{fig::cdf.ITT.stocks} show the analogous distributions for the unnormalized time series $\delta t_i$. Here the cdf tails perform significantly different than in the main plots, but this may be considered as a misleading effect due to the significantly different values of mean $\langle \delta t \rangle$ and standard deviation $\sigma_{\delta t}$ of the considered ITT time series. Note that the unsigned character of $\delta t_i$ implies the cdf tail behaviour can influence not only $\sigma_{\delta t}$ but also $\langle \delta t_i \rangle$, making both quantities effectively related. This is why the German stocks with significantly higher $\langle \delta t \rangle$ - KAR and LIN - have a shorter distribution tail measured in $\sigma_{\delta t}$. On the other hand, when we calculate the distribution in original time units without normalization (insets), the stocks with large $\langle \delta t \rangle$ have the longest distribution tail and stocks with small $\langle \delta t \rangle$ have the shortest distribution tail.

A similar analysis for the cryptocurrency ITTs leads to the results that are displayed in Fig,~\ref{fig::cdf.ITT.crypto}. For small multiplicities of $\sigma_{\delta t}$, the empirical cdfs can in most cases be fitted with the SE model. However, if compared with the results for the stocks, this model performs on average much worse with only a few cases, in which the SE distribution follows its empirical counterpart up to more than $10\sigma_{\delta t}$. For all the data sets, the exponential model does not agree with the data either, even though for the time series representing XRP and LTC from Kraken it is relatively the closest one (bottom panels). The cryptocurrencies on less liquid platforms with higher $\langle \delta t \rangle$ (Kraken, Bitstamp, and HitBTC - see Tab~\ref{tab::data.statistics}) have shorter distribution tails measured in $\sigma_{\delta t}$ and the longest tails for unnormalized time series $\delta t$ distributions. The opposite situation, the longest distribution tail measured in $\sigma_{\delta t}$ and the shortest one for the unnormalized $\delta t$, can be observed for the most liquid exchange -- Binance.

\begin{figure}
\includegraphics[width=0.45\textwidth]{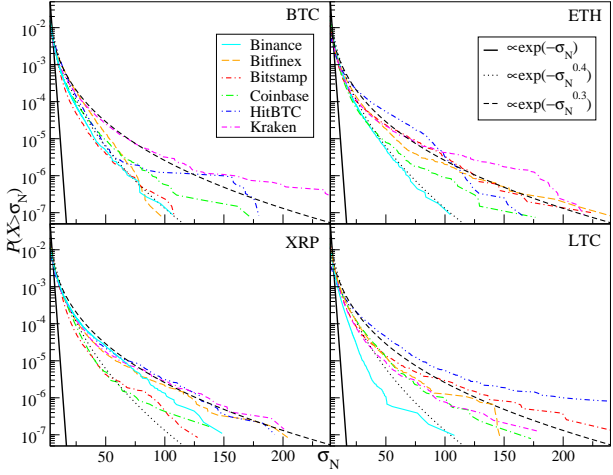}
\caption{Cumulative distribution functions of the normalized number of transactions $\hat{N}_{\Delta t,i}$ in units of standard deviation $\sigma_{N_{\Delta t}}$ with $\Delta t=10$s for the cryptocurrencies: BTC (top left), ETH (top right), XRP (bottom left), and LTC (bottom right). In each plot cdfs for the data from six trading platforms are shown, denoted as in Fig.~\ref{fig::cdf.ITT.crypto}. Different variants of the SE model with $\alpha=0.3$ (dashed), $\alpha=0.4$ (dotted), and $\alpha=1.0$ (solid) serve as guide for an eye.}
\label{fig::cdf.number.semilog.crypto}
\end{figure}

Another quantity that we shall attempt to model is the normalized number of transactions $\hat{N}_{\Delta t,i}$ realised in intervals of $\Delta t=10$s. These time series much more sparsely than $\hat{\delta t_i}$ show values that equal 0. Fig.~\ref{fig::cdf.number.semilog.crypto} documents that in this case no empirical cdf obeys the exponential distributions, while the SE distributions are able to account for for the empirical results in some cases (Kraken up to $80\sigma_{N_{\Delta t}}$ for BTC; HitBTC up to $50\sigma_{N_{\Delta t}}$ for BTC; Bitfinex, HitBTC, and Kraken up to $200\sigma_{N_{\Delta t}}$ for XRP). A particularly interesting is the result for Binance, where the cdfs obtained from BTC, ETH, and XRP follow the stretched exponentials over the whole range of time series values. In contrast, LTC does not exhibit such an effect at all.

\begin{figure}
\includegraphics[width=0.45\textwidth]{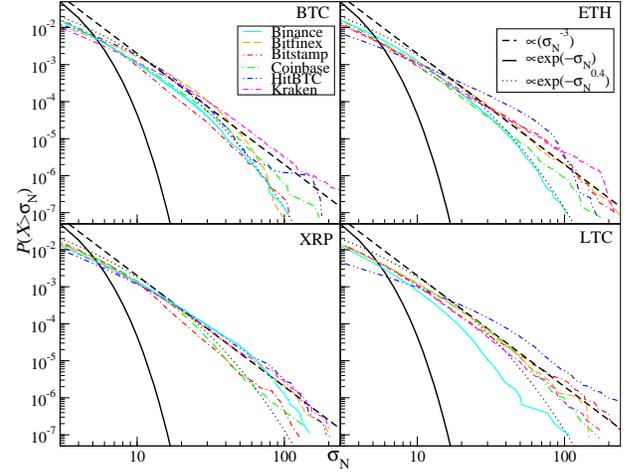}
\caption{The same cumulative distribution functions as in Fig.~\ref{fig::cdf.number.semilog.crypto} plotted in double logarithmic scale. The SE model with $\alpha=0.4$ (dotted), the exponential model ($\alpha=1.0$, solid), and the power-law model with tail exponent $\beta=3$ (dashed) serve as guide for an eye.}
\label{fig::cdf.number.loglog.crypto}
\end{figure}

Another model that has to be tested against empirical data is the power-law distribution with a tail exponent $\beta=3$ (the inverse cubic power law dependence~\cite{GopikrishnanP-1999a,PlerouV-1999a}). Fig.~\ref{fig::cdf.number.loglog.crypto} shows essentially the same empirical distributions as Fig.~\ref{fig::cdf.number.semilog.crypto}, but plotted in double logarithmic scale in order to simplify detection of possible power-law scaling. In the case of BTC, we see two clear examples of a partial agreement between the data and the power-law model: Bitstamp for medium time series values ($\beta > 3$) and Kraken for the whole range of $N_{\Delta t,i}$ (for $\beta\approx 3$). HitBTC can be better fitted by the SE distribution with $\alpha=0.4$ In the case of ETH, more data sets show power-law scaling for at least part of time series value range with $\alpha < 3$. The exception is HitBTC, whose cdf does not exhibit any scaling region. Such an exception is Coinbase, which is more SE-like, in the case of XRP, which generally shows the inverse cubic power-law dependence. In parallel, the exception for LTC is also HitBTC, which in this case is not compatible with any of the models.

\begin{figure}
\includegraphics[width=0.45\textwidth]{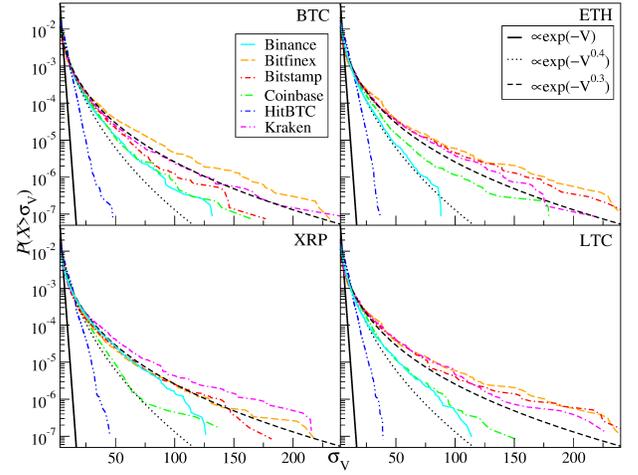}
\caption{Cumulative distribution functions of the normalized volume $\hat{V}_{\Delta t,i}$ in units of standard deviation $\sigma_{V_{\Delta t}}$ with $\Delta t=10$s for the cryptocurrencies: BTC (top left), ETH (top right), XRP (bottom left), and LTC (bottom right). In each plot cdfs for the data from six trading platforms are shown, denoted as in Fig.~\ref{fig::cdf.ITT.crypto}. Different variants of the SE model with $\alpha=0.3$ (dashed), $\alpha=0.4$ (dotted), and $\alpha=1.0$ (solid) serve as guide for an eye.}
\label{fig::cdf.volume.semilog.crypto}
\end{figure}

\begin{figure}
\includegraphics[width=0.45\textwidth]{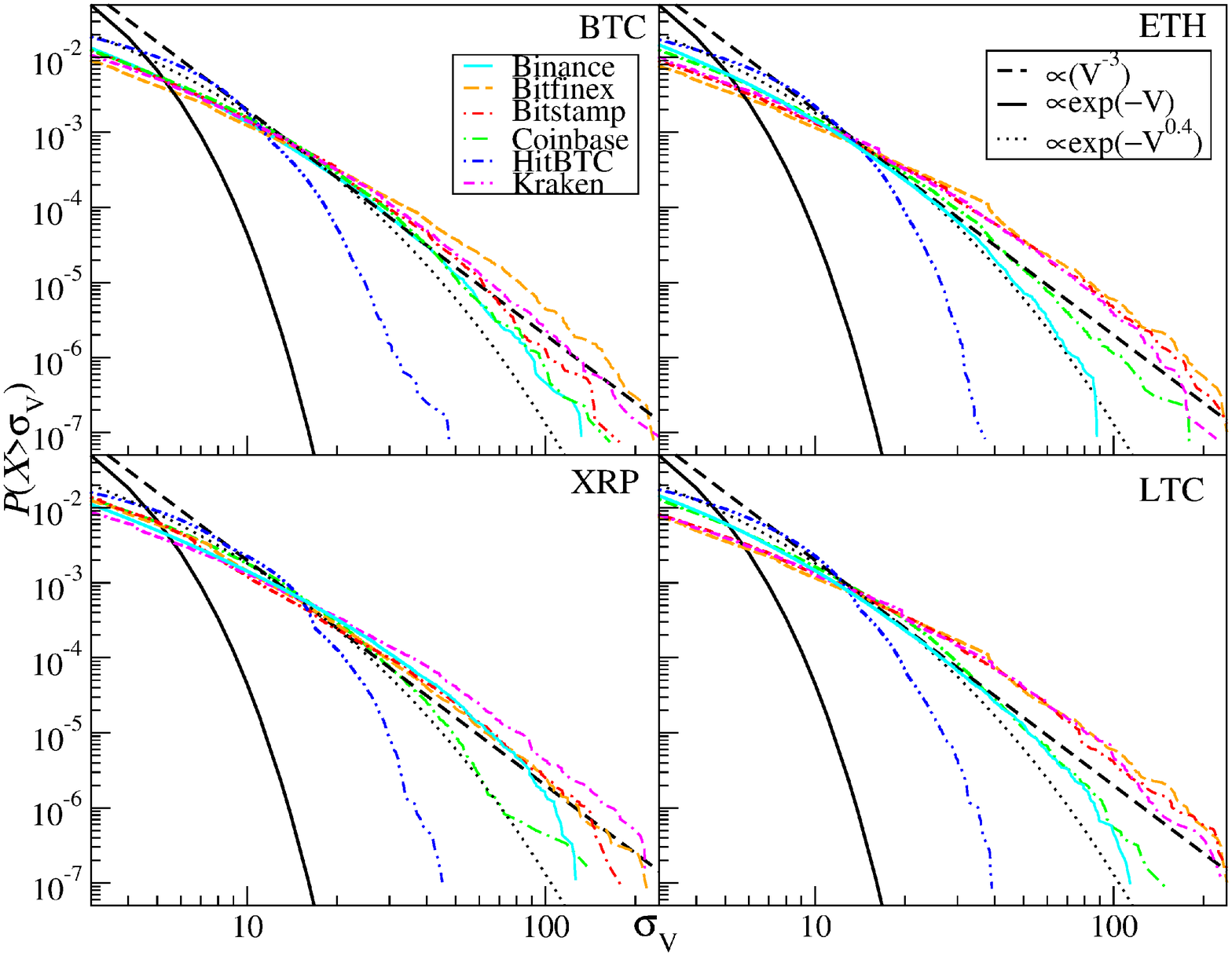}
\caption{The same cumulative distribution functions as in Fig.~\ref{fig::cdf.volume.semilog.crypto} plotted in double logarithmic scale. The SE model with $\alpha=0.4$ (dotted), the exponential model ($\alpha=1.0$, solid), and the power-law model with tail exponent $\beta=3$ (dashed) serve as guide for an eye.}
\label{fig::cdf.volume.loglog.crypto}
\end{figure}

Fig.~\ref{fig::cdf.volume.semilog.crypto} presents the cdfs calculated from the time series of normalized volume traded $\hat{V}_{\Delta t,i}$. Here the SE model seems to be in agreement with the data for BTC, ETH, and LTC even for the HitBTC platform, even though it is close to an exponential distribution. Relatively the worst compatibility between the model and the data is seen for XRP, but the corresponding disagreement is predominantly connected with the largest values of the time series, while the smaller values are also well fitted by the SE distribution. As the traded volume cdf behaves as the SE function, an attempt to model its tails with the power-law functions (Fig.~\ref{fig::cdf.volume.loglog.crypto}) largely fails. A trace of scaling can be found only in few cases: BTC from Kraken and Coinbase,  ETH from Bitfinex and Coinbase, and XRP from Bitfinex, Bitstamp, and Kraken. No compliance with this model can be found for LTC. We can also distinguish here an unusual behaviour of the HitBTC platform, where the volume distribution has the shortest tails.

\section{Conclusions}

In our work we studied high-frequency, tick-by-tick data from a few cryptocurrency trading platforms, including the largest one -- Binance. Unlike our earlier analysis~\cite{WatorekM-2021a}, which was devoted to price fluctuations, here we focused on the statistical properties of the inter-transaction times, the number of transactions in time unit, and volume for 4 major cryptocurrencies: BTC, ETH, XRP, and LTC.

All these quantities show significant variation with time related both to changes that the whole cryptocurrency market (or even the whole economy) has experienced during several years: the speculative bubbles and subsequent trend reversals, the outburst of pandemics, the progressive development of the cryptocurrency market with more and more investors focusing their attention there with resultant improved liquidity, the emergence of new trading platforms and new cryptocurrencies that reshape the market internally, and so on. Such a long-term variation of the market properties contributes to the already existing long-range correlations in the quantities characterising the market and the individual assets. These correlations, which are often power-law decreasing, are in turn responsible for the fractal structure of the market time series. Our analysis confirms the earlier published conclusions that the inter-transaction times for the stock market data are multifractal and, for the first time, gives evidence that the same is true for the cryptocurrency market. Our study shows also that the ITTs display a right-side asymmetry of the singularity spectra, which indicates that mainly the small values of ITTs are responsible for the multifractality, while the large values show poorer multifractal behaviour. As small ITTs correspond to the intensive trading activity, our results show that on the cryptocurrency market shows richer multifractality during its most active periods.

Another observation we made is that the highest level of the detrended cross-correlations for the number of transactions and volume, which can be viewed as natural, while the relatively lowest yet still significant for the number of transactions and volatility. We found that, generally, the inter-transaction times of the cryptocurrency trading show significantly worse agreement with the stretched-exponential model compared to the ITTs of large-capitalization stocks from the American and German markets.  On the other hand, if we look at individual time series from different platforms, in some cases the SE model cannot be discarded. The number of transactions in time unit, which in our case was 10s, displays mixed evidence: there are data sets that seem to be in better agreement with the power-law model and there are also ones whose cdf tails can be well-fitted by the stretched exponentials.

An interesting observation is that the results can differ even among the platforms: the number of transactions for the same cryptocurrency can be modelled by the SE distribution on one platform, while it cannot on another platform, where it is more compatible with the power-law model. This is not the only quantity that shows such duality as similar observations can be made for the time series of traded volume (which cannot come as a surprise, because both quantities are strongly cross-correlated). This discrepancy of the statistical properties of the otherwise parallel data representing different trading platforms can serve as an intriguing topic for future research. This is especially true for the platforms where data reveal even more peculiar properties, like in the case of the HitBTC platform. As some platforms face strong criticism regarding some practices, like wash trading, they allegedly have been carrying out, an in-depth analysis of the related data might give answers to such criticisms. Our results can be considered a clue as to where to search for such answers.

%

\section*{Data Availability Statement}

\noindent
The data that support the findings of this study are openly available from the trading platforms cited in References.

\nocite{*}
\bibliography{aipsamp}

\end{document}